\documentstyle[11pt]{article}
\def\build#1_#2^#3{\mathrel{
\mathop{\kern0pt#1}\limits_{#2}^{#3}}}
\def\la{\mathrel{\mathpalette\fun <}}

\def\fun#1#2{\lower3.6pt\vbox{\baselineskip0pt\lineskip.9pt
        \ialign{$\mathsurround=0pt#1\hfill##\hfil$\crcr#2\crcr\sim\crcr}}}

\def\ben{\begin{equation}}
\def\be#1{\begin{equation}\label{eq:#1}}
\def\ee{\end{equation}}

\begin{document}

\begin{center}
\vspace{.2in}

\baselineskip 24pt

{\Large \bf Microlensing in a Prolate All-Macho Halo}
\vspace{.2in}

Gilbert P. Holder and Lawrence M. Widrow \\
\vspace{.1in}
\medskip
$^1${\it Department of Physics\\
Queen's University, Kingston, Canada, K7L 3N6}\
\end{center}

\vspace{.3in}

\pagestyle{plain}

\begin{abstract}

It is widely believed that dark matter halos are flattened, that is
closer to oblate than prolate.  The evidence cited is based largely on
observations of galaxies which do not look anything like our own and
on numerical simulations which use {\it ad hoc} initial conditions.
Given what we believe to be a ``reasonable doubt'' concerning the
shape of dark Galactic halo we calculate the optical depth and event
rate for microlensing of stars in the LMC assuming a wide range of
models that include both prolate and oblate halos.  We find, in
agreement with previous analysis, that the optical depth for a
spherical (E0) halo and for an oblate (E6) halo are roughly the same,
essentially because two competing effects cancel approximately.
However the optical depth for an E6 prolate halo is reduced by $\sim
35\%$.  This means that an all-Macho prolate halo with reasonable
parameters for the Galaxy is consistent with the published
microlensing event rate.

\end{abstract}

\centerline{\bf 1.~Introduction}
\bigskip

The announcements by the MACHO (Alcock et al.\,1993) and EROS (Aubourg
et al.\,1993) collaborations of candidate microlensing events towards
the Large Magellanic Cloud (LMC) generated tremendous excitement by
suggesting that the identity of one contribution to the dark Galactic
halo had been found.  Yet despite two years of additional observations
and intense analysis the implications of the microlensing results
remain unclear.  The event rate toward the LMC, as determined by the
MACHO collaboration for example, is significantly below what one would
expect if Machos are the dominant component of the ``standard''
Galactic halo but significantly above what one would expect if the
lenses are from known stellar populations.  The best guess is that
5\%-30\% of the halo is in Machos.  An all-Macho halo cannot be ruled
out though this would require a model for the Galaxy that is only
marginally acceptable (Alcock et al.\,1995a,b, Gates, Gyuk, \& Turner
1995a,b).

Attempts to derive an accurate and robust estimate for the mass
density of Machos in our Galaxy are hampered by statistical
uncertainties due to the small number of events and systematic
uncertainties in the detection efficiency for the experiments.  But
even if these difficulties could be overcome we would be left with
important uncertainties in modelling our Galaxy.  In particular the
parameters which describe our Galaxy, such as the local rotation speed
and the rotation speed far from the Galactic center, are
not well constrained.

Perhaps the most important unknown in Galactic models is the shape of
the halo.  The conventional wisdom is that halos are flattened.  With
this in mind a number of researchers have calculated the optical
depth for oblate versus spherical model halos (Sackett \& Gould 1993;
Frieman \& Scoccimarro 1994; Alcock et al.\,1994a; Gates, Gyuk \&
Turner 1995a,b).  The optical depth to microlensing toward the LMC is
found to be roughly the same for an E6 (axial ratio $q=0.4$) oblate
halo and an E0 ($q=1.0$) spherical halo.  (Following standard notation
$a>b>c$ refer to the three semi-axis lengths for a triaxial halo.  For
a spheroidal halo, two of these lengths are equal and the shape of the
halo is specified by a single parameter $q$ defined to be the ratio of
the symmetry axis to the equatorial axis.) This is essentially because
two competing effects, one geometric and the other related to the
central density, cancel approximately.  The general conclusion is that
the total mass in Machos is constrained tightly by the microlensing
data while the mass fraction is model-dependent (Alcock et
al.\,1995a,b; Gates, Gyuk, \& Turner 1995b). (This conclusion is based
on analysis in which other halo parameters such as the thickness of
the disk are also varied.)

Here we consider a broad spectrum of models ranging from E6 prolate
($q=2.5$) to E6 oblate.  In each model the symmetry axis is chosen to
be perpendicular to the disk.  For a fixed rotation curve, the central
density decreases as we increase $q$ and this leads to a decrease in
the optical depth.  The optical depth for an E6 prolate halo, for
example, is reduced relative to the optical depth for either an E6
oblate or spherical halo by $\sim 35\%$.  This is enough to change
significantly the conclusions based on the MACHO data.  In particular
an all-Macho halo with reasonable halo parameters is now allowed.
Moreover our models provide an example where the total mass in Machos
inferred from the data is different from that found assuming either an
oblate or spherical halo.

Prolate halos are unconventional to say the least.  The overriding
view of researchers in the field is that halos are spherical ($a\sim
b\sim c$), oblate ($a\sim b> c$), or triaxial ($a>b>c$) (Rix 1995,
Sackett 1995).  Evidence for the shape of the dark Galactic halo comes
mainly from dynamical modelling of metal-poor halo stars.
Observations constrain $q_s$, the axial ratio for the (metal-poor)
stellar halo, to be $0.6\la q_s\la 0.8$ (Gilmore, Wyse, Kuijken 1989).  The
simplest assumption, that
the dark Galactic halo has similar kinematical properties, implies
$q<1$, i.e., an oblate dark Galactic halo.  However, the relation
between $q$ and $q_s$ may not be so simple.  van der Marel (1991) has
derived this relation under the assumption that the halo stars are in
hydrostatic equilibrium (i.e., obey the Jeans equations).  The results
indicate that this relation is extremely sensitive to assumptions made
about the local velocity ellipsoid of the halo stars.  Indeed the
axial ratio for the dark matter can be much larger than that for the
stellar halo and in some cases greater than 1.

There is compelling evidence from observations of other galaxies that
oblate halos exist. Observations of rare polar-ring galaxies
(eg. Sackett et al. 1994) find flattened halos with $0.3\la c/a\la
0.6$. As well observations of X-ray emitting gas around ellipticals,
used to trace the gravitational potential of the halos around these
galaxies, also find flattened halos (Franx, van Gorkum \& de Zeeuw
1994). However, there is no reason to expect all dark halos to have
the same shape especially when the visible components of galaxies are
so different.  Of particular interest would be observations of spiral
galaxies such as our own.  At present the only such observations are
of flaring HI gas layers (Olling 1995). The most detailed observations
of this type are for the spiral galaxy NGC 4244 with the result $0.2\la
c/a \la 0.8$. Clearly more observations of spirals are be needed
before any definite conclusions are to be drawn.

Theoretical evidence in favour of oblate halos comes largely from
numerical simulations.  In fact the halos found in dissipationless
simulations are triaxial with nearly prolate halos being more common
than oblate ones.  One must consider the effects of dissipational
matter in order to see why oblate halos are favoured.  Katz \& Gunn
(1991) simulate the gravitational collapse of constant density
spherical perturbations consisting of both dark matter and gas.  These
are isolated perturbations (i.e., no external tidal effects) and so
angular momentum must be put in by hand.  This is done by imposing
solid-body rotation on the initial perturbations.  The simulations show
that halos that would be triaxial or prolate in the absence of a gas
component are nearly oblate once the gas is included.  However it is
not known whether these results are due to the very special initial
conditions used and in particular whether the conclusions would be
different if the rotation axes for the gas and dark matter were
misaligned initially.  Indeed the gas and dark matter may have
collapsed at different times therefore experiencing different tidal
torques.  As an interesting aside, Katz \& Gunn (1991) find that the angular
momentum axes often become misaligned through the course of
simulations as angular momentum is transferred from the dark matter to
the disk.

Dubinski (1994) models the effects of dissipational matter by growing
a disk in the center of a dark halo that is the output from a
cosmological N-body simulation.  The symmetry axis of the disk is
aligned with the short axis of the halo which in turn coincides
roughly with the rotation axis of the halo.  The result, that an
initially prolate-triaxial halo evolves into an oblate-triaxial one,
is not unexpected given the set-up for the simulation.  But again we
are lead to wonder whether the conclusions would be different if the
rotation axis were misaligned initially.

Clearly, there is a need for numerical simulations that model
self-consistently both the dark matter and the gas with enough
resolution to study individual galaxies and large-scale effects. Steps
in this direction have been made by Evrard, Summers \& Davis
(1994). While the small-scale resolution is not fantastic their
results do suggest that halos become rounder in the presence of
dissipational matter.  But in the end the halos are still, on average,
more prolate than oblate.

Finally it is worth pointing out that all of the simulations described
above assume that galaxies form in a hierarchical clustering scenario
such as the Cold Dark Matter model.  Clearly if Machos are the
Galactic dark matter then these ideas would require serious revision.

\bigskip\bigskip

\centerline{\bf 2.~Optical Depth to Microlensing from Prolate Halos}

\bigskip\bigskip

The dark Galactic halo is modelled as a cored isothermal spheroid (see,
e.g., Sackett \& Gould 1993).  The density profile is given by

\be{density}
\rho(R,z)=
{{v_{\infty}^2} \over {4 \pi G}}\, {1 \over {R_c^2 + R^2+{z^2 /
q^2}}} \,f(q)
\ee

\noindent where $z$ is the distance from the equatorial plane, $R$ is the
distance from the $z$ axis, $R_c$ is the core radius, and

\be{deff}
f(q)=\left\{
	\begin{array}{ll}
\frac{\sqrt{1-q^2}}{q\arccos{q}}&\mbox{$q<1$} \\
\\
1 & \mbox{$q=0$}\\
\\
\frac{\sqrt{q^2-1}}{q\cosh^{-1}{q}}&\mbox{$q>1$}
	\end{array}
	\right.
\ee

\noindent $v_\infty$ is the asymptotic circular speed due to the
halo alone.  The tightest constraints on $v_\infty$ come from the
local circular speed $v_c$.  $v_c$ receives nontrivial contributions
from the disk and bulge and therefore $v_\infty$ is model dependent.
We use $v_\infty=200\,{\rm km/sec}$ for the standard Galactic model.
Lower values for $v_\infty$ require a heavier disk and/or bulge
but are still allowed by the observations.

The optical depth $\tau_{\rm LMC}$ is defined as the probability that
the light from a star in the LMC will be amplified via gravitational
microlensing by a factor $A\ge1.34$.  For this to occur, a Macho must
pass within a distance $R_E=\left (4GML(D-L)/c^2D\right )^{1/2}$ of
the line of sight to the star (eg. Paczynski 1986, Griest 1991).  Here $D$ and
$L$ are the distance to
the background star and Macho respectively, $M$ is the mass of the
Macho, and $R_E$ is the Einstein radius.  The optical depth is given
by

\be{od}
\tau_{\rm LMC}=\int_0^D dL {4\pi GL(D-L) \over
c^2\;D}\rho(L)
\ee

\noindent where $\rho(L)$ is the mass density in Machos along the lines
of sight to the LMC.  The uncertainties in the optical depths allowed
by the MACHO group's data are difficult to pin down, but following
Gates, Gyuk \& Turner 1995, we take $2.0\times 10^{-7}$ as a $2\sigma$
upper bound (95$\%$ confidence level) of $\tau_{\rm LMC}$.

Figure 1 gives $\tau_{\rm LMC}$ as a function of $q$ for $R_c=5\,{\rm
kpc}$ and three values of $v_\infty$, with the sun taken to be $8.5\,
{\rm kpc}$ from the galactic center. The physical effects at work are
easy to understand. As the halo becomes more oblate the constraint
that the rotation curve remain largely unchanged requires an increase
in the density of the halo by the factor $f(q)$. This tends to
increase the optical depth. At the same time our line of sight to the
LMC is passing through less of the halo and so there is a decrease in
the optical depth. These two factors lead to the turnover in
$\tau_{\rm LMC}$ with the maximum occurring for $q\simeq 0.4$
depending on the value of $v_\infty$.  No ``reasonable model''
($0.4\la q\la 2.5$) is consistent with $v_\infty=200\,{\rm km/sec}$.
However an all-Macho E6 prolate halo can fit the observations with
$v_\infty=160\, {\rm km/s}$.  This value for $v_\infty$ is comfortably
allowed by observations of our Galaxy provided one assumes appropriate
masses for the for disk and bulge.  Note that if we limit ourselves to
oblate halos then we must choose $v_\infty\la 135 {\rm km/sec}$ which
is extremely low though not entirely ruled out.

The MACHO group (Alcock et al. 1995a,b) argue that the LMC microlensing
results provide a model-independent constraint on the mass in Machos
within $50\;{\rm kpc}$ of the Galactic center.  To be precise, it is
$v_\infty$ that is constrained. The small variance in the optical
depth across the range of oblate models means that the only way to
reduce the optical depth is to reduce $v_\infty$. That is, for a given optical
depth, there is a very small spread in $v_\infty$ that is
allowed. This is what is meant by a model-independent constraint on
the circular speed due to Machos. However, if the range of models is
extended to include prolate models, then this constraint is no longer
model-independent. For a given optical depth, there can now be a range
of flattenings allowed corresponding to different models of the Galaxy
each with a different $v_\infty$,

\bigskip\bigskip

\centerline{\bf 3.~Expected Number of Events}

\bigskip\bigskip

It has been pointed out (Griest 1991) that $N_{\rm exp}$, the expected number
of
events for their experiment, is more important than the optical depth
when comparing predictions with observations.  $N_{\rm exp}$ is given
by

\ben
N_{\rm exp}=E\int_0^\infty
{d\Gamma \over d\hat{t}}
\;\epsilon(\hat{t})\;d\hat{t},
\ee

\noindent where $\hat{t}$ is the event duration, $\epsilon(\hat{t})$ is the
detection efficiency, and $E$ is the total ``exposure" given in units
of ``star yr".  ${d\Gamma / d\hat{t}}$ is the differential event rate
where the total event rate $\Gamma$ has units ``events/star/yr".
$\Gamma$ depends on the velocity distribution and mass function of the
Machos as well as their mass density.  For simplicity we assume that
all of the Machos have the same mass $M$.  In addition we assume a
Maxwellian velocity distribution centered on $v_c=200\,{\rm km/sec}$
so that the distribution function $f({\bf r},\,{\bf
v})\propto\rho({\bf r})\exp{\left (-v^2/v_c^2\right )}$.  We then have (Griest
1991, Alcock et al. 1995b)

\ben
{d\Gamma \over dt} = {32 \over {\hat{t}^4 m
v_c^2}} \int_0^D \rho(L) R_E^4(L) \;\exp \! \left(-{4 R_E^2(L) \over
\hat{t}^2 v_c^2} \right) dL .
\ee

\noindent Figure 2 is a plot of $N_{\rm exp}$ as a function $M$ for three
models: E6 oblate, E0, and E6 prolate, using this Maxwellian
distribution.  A rough fit to the published
expected efficiencies of the MACHO collaboration (Alcock et al 1995a) is used
for $\epsilon(\hat{t})$
and all other parameters are the same as for Figure 1. It can be seen
that the oblate(E6) and spherical models predict nearly identical
event rates, while the predicted event rate for a prolate (E6) halo is
substantially reduced. More importantly, the excluded range
(i.e. $N_{\rm exp}\ge 7.7$) has been reduced such that an all-Macho
halo consisting of brown dwarfs is not ruled out.

The Maxwellian distribution described above does not satisfy the
Vlasov equation and so technically cannot describe a realistic galaxy.
Recently several authors have attempted to address this problem by
using the so-called power-law models for the Machos.  When taken alone
these model halos, which are constructed from simple power-law
functions of the energy and angular momentum, are guaranteed to be
solutions of the Vlasov equation.  This is no longer the case when the
other components of the galaxy are included. It has been pointed out
that at a distance halfway to the LMC the disk contributes less than
$4\%$ to the circular velocity (Evans \& Jijina 1994). However
statistically speaking well over half of the lensing events are likely
to occur at a closer distance (eg. Kan-Ya et al. 1995, Roulet et al 1994).  The
$4\%$
figure therefore represents a lower limit to the error incurred in
neglecting the disk. The net effect is to underestimate systematically
the velocities of the Machos. This leads to a lower than expected
event rate along with a shift towards longer timescales.  Figure 12 of
Alcock et al 1995b  shows $d\Gamma/d\hat{t}$ for the standard halo
with a Maxwellian velocity distribution along with the corresponding
spherical power-law halo. The power-law halo has a slightly lower event rate
along with a shift towards longer timescales as expected from the
discussion above. For an oblate halo the lenses are, on average,
closer to us than for the spherical case and so the error that comes
from neglecting the disk is even greater.  This explains
quantitatively Figure 2 of Evans \& Jijina (1994) where the lensing
rate for an E6 halo appears to be slightly lower and shifted to a
longer timescale as compared to a spherical halo.

The great advantage of the power-law halo models is that the distribution
functions are analytic making lensing calculations relatively simple.
There are however indications that they may not correspond to realistic
halos.  In particular, while the equipotential surfaces are spheroidal,
the density profiles are dimpled at the poles for $q \la 0.85$ (Evans
1993). Models which correctly take into account the gravitational
couplings between disk, bulge, and halo have recently been constructed
(Kuijken \& Dubinski 1995).  In addition these models have more reasonable
density profiles.  The models are not analytic and so lensing calculations
are more complicated (though still straightforward).  Moreover the
range in $q$ for these models is fairly limited and only includes
a small fraction ($q\la 1.1$) of the prolate halos.  It is for these
reasons that we have taken the simplest approach wherein velocities
are Maxwellian.

\bigskip\bigskip

\centerline{\bf 4.~Conclusions and Discussion}

\bigskip\bigskip

Our primary result is that the conclusions one draws from the
published data on microlensing towards the LMC depend sensitively on
the model one chooses for the Galactic halo.  In particular, an
all-Macho prolate halo is still consistent with the data.  Conversely,
a spherical or oblate all-Macho halo is all but ruled out.  Few
theorists or observers believe that halos are prolate.  However a
careful review of the evidence cited in favour of flattened halos
suggests that a prolate spheroid is still a viable model for our
Galaxy's halo.

It should be stressed that the reduction in the predicted microlensing
rates for model prolate halos is independent of the shape of the
rotation curve.  That is, for a given rotation curve, an oblate or
spherical model halo will have a higher density of Machos and
therefore a higher predicted microlensing event rate.  Even if the shape of the
Galactic
rotation curve were known precisely, the fraction of dark matter
composed of Machos, as determined from experiments like MACHO and
EROS, would remain uncertain until more information about the shape of our halo
could be obtained.

\bigskip\bigskip

\centerline{\bf Acknowledgements}

\bigskip

It is a pleasure to thank  W-Y Chau, J. Dubinski, G. Evrard, P. Sackett, and
J. Sellwood for useful conversations.  This work was supported in part
by a grant from the Natural Sciences and Engineering Research Council
of Canada.

\bigskip\bigskip

\centerline{\bf References}

\bigskip\bigskip

\noindent Alcock, C. {\it et al.} 1993, Nature, 365, 621

\noindent Alcock, C. {\it et al.} 1995a, ApJ, 449, 28

\noindent Alcock, C. {\it et al.} 1995b, astro-ph 9506113

\noindent Aubourg, E. {\it et al.} 1993, Nature, 365, 623

\noindent Dubinski, J. 1994, ApJ, 431, 617

\noindent Evans, N.W. 1993, MNRAS, 260,191

\noindent Evans, N.W.,\& Jijina, J. 1994, MNRAS, 267, L21

\noindent Evrard, A. E., Summers, F. J., \& Davis, M. 1994, ApJ, 422, 11

\noindent Franx, M., van Gorkom, J. H., \& de Zeeuw, T. 1994, 436, 642

\noindent Frieman, J. A. \& Scoccimarro, R. 1994, ApJ, 431, L23

\noindent Gates, E., Gyuk, G.,\& Turner, M. S. 1995a, ApJ, 449, L123

\noindent Gates, E., Gyuk, G.,\& Turner, M. S. 1995b, astro-ph/9508071

\noindent Gilmore, G., Wyse, R. F. G.,\& Kuijken, K. 1989, ARAA, 27, 555

\noindent Griest, K. 1991, ApJ, 366, 412

\noindent Kan-ya, Y., Nishi, R.,\& Nakamura, T. 1995, astro-ph/9505130

\noindent Katz, N. \& Gunn, J. E. 1991, ApJ, 377, 365

\noindent Kuijken, K. \& Dubinski, J. 1995, MNRAS, in press

\noindent Olling, R. \& van Gorkom, J. H. 1995, in ``Dark Matter'', eds. S.S.
Holt \& C.L.  Bennett (New \indent York:AIP) p121

\noindent Paczynski, B. 1986, ApJ, 304, 1

\noindent Rix, H.-W. 1995, astro-ph/9501068

\noindent Roulet, E., Mollerach, S., Giudice, G. F. 1994, astro-ph/9405041

\noindent Sackett, P. D. \& Gould, A. 1994, ApJ, 419, 648

\noindent Sackett, P. D. 1995, astro-ph/9508098

\noindent Sackett, P. D., Rix, H., Jarvis, B. J., Freeman, K. C. 1994, ApJ,
436, 629

\noindent van der Marel, R. P. 1991, MNRAS, 248, 515

\bigskip\bigskip

\centerline{\bf Figure Captions}

\bigskip\bigskip

\noindent {\bf Figure 1}: Plot of optical depth to the LMC, $\tau_{LMC}$ vs.
flattening parameter q for $v_\infty$=200,160,135 km/s. The solid line at
$\tau_{LMC}=2\times 10^{-7}$ represents a $2\sigma$ upper bound from
observations.

\bigskip

\noindent {\bf Figure 2}: Plot of expected number of events, $N_{exp}$ vs. mass
of lenses for three halo models: E0, E6 oblate and E6 prolate.

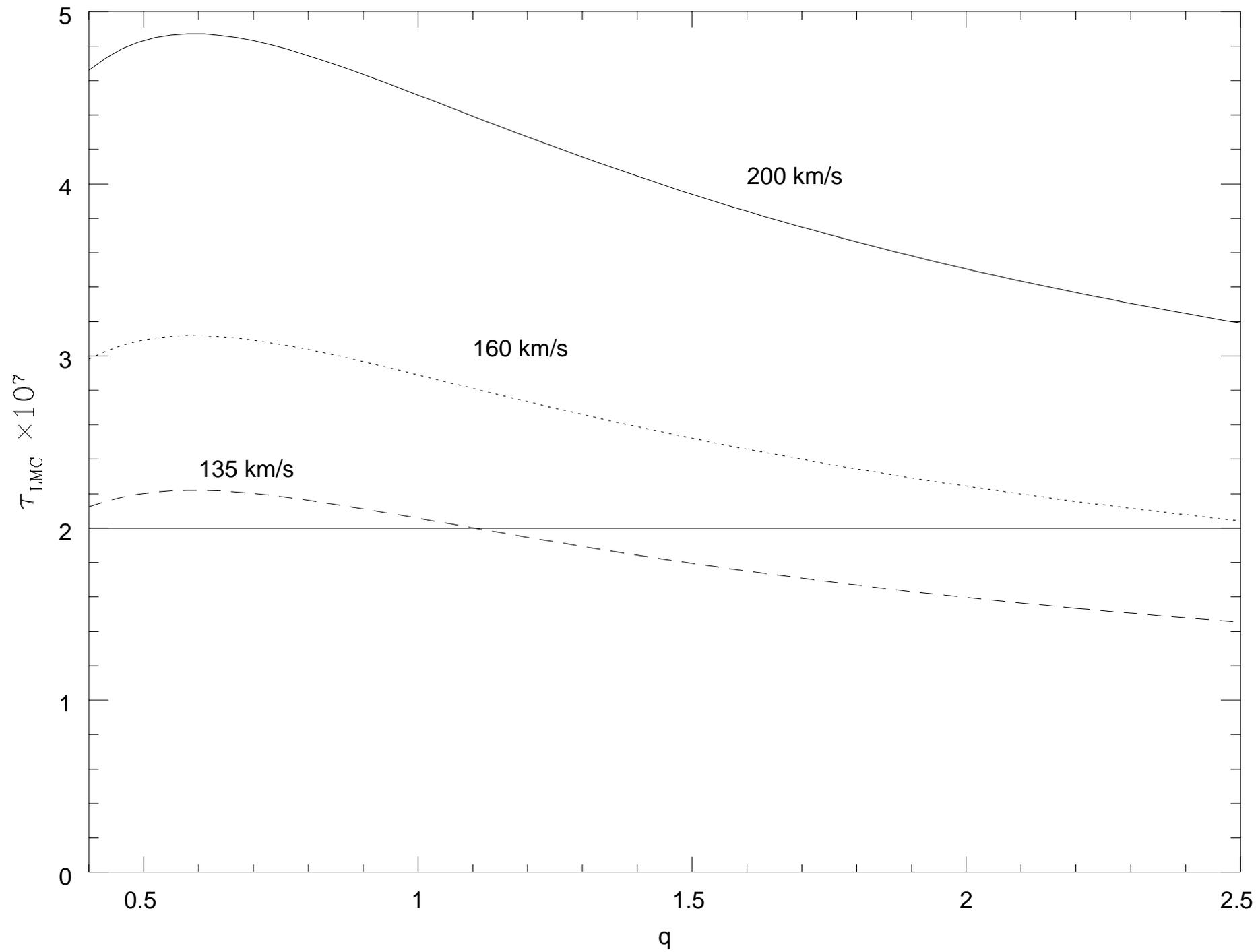
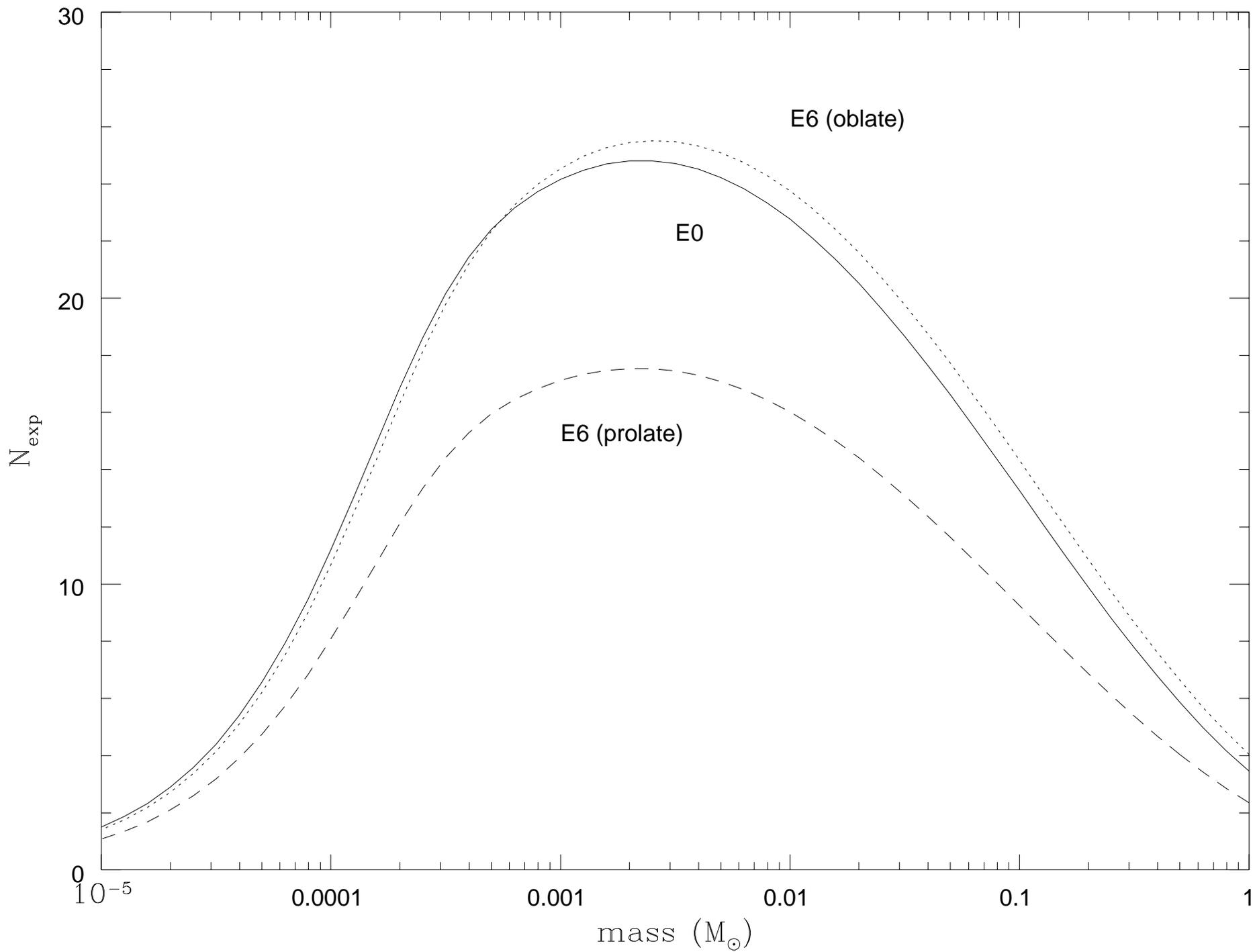
\end{document}